# Bandgap engineering of $Cd_{1-x}Sr_xO$

I. Khan, Iftikhar Ahmad[*], B. Amin, G. Murtaza, Z. Ali

Materials Modeling Lab, Department of Physics, Hazara University, Mansehra, Pakistan

*Corresponding author email address: ahma5532@gmail.com

**Abstract**

Structural, electronic and optical properties of $Cd_{1-x}Sr_xO$ ($0 \leq x \leq 1$) are calculated for the first time using density functional theory. Our results show that these properties are strongly dependent on *x*. The space group of the compound changes from Fm-3m (x = 0) to Pm-3m (x = 0.25) to P4/mmm (x = 0.50) to Pm-3m (x = 0.75) and finally to Fm-3m (x = 1). The linear relationship of lattice constant with concentration x confirms that the material obeys Vegard's law except for x = 0.50 which is tetragonal rather than cubic. The nature of bond of the compound also changes from covalent to ionic as Sr increases from 0 to 100%. It is also found that $Cd_{1-x}Sr_xO$ is an indirect bandgap compound for the entire range of x. While the bandgap of the alloys increases from 0.85 to 6.00 eV with the increase in Sr concentration. Frequency dependent dielectric functions $\varepsilon_1(\omega)$, $\varepsilon_2(\omega)$, refractive index $n(\omega)$ and absorption coefficient $\alpha(\omega)$ are also calculated. It is also found that $Cd_{0.50}Sr_{0.50}O$ is an anisotropic material. The larger value of the extraordinary refractive index confirms that the material is positive birefringence crystal. The peak value of refractive indices shifts to higher energy regions with the increase in Sr.

# I. Introduction

Group II-VI semiconductors have extensively studied due to their effective use in optoelectronic industry. These compounds are commonly used in many established commercial electronic and optoelectronic devices operating in blue to ultraviolet spectral regions [1-6]. Optically transparent and electrically conductive nature of CdO makes it an important member of the group. Its industrial applications are not limited to solar cells, transparent electrodes, ohmic contacts to LEDs, smart windows, optical communications, flate panel displays, photo-transistors, photovoltaics, gas sensors, low-emissive windows, thin-film resistors [7-13] but it is also efficiently used as a high reflective compound in the infrared region while high transparent material in the visible region [11]. Bandgap of CdO is 0.85 eV, while strontia (SrO) is an insulator with bandgap of 5.71 eV [14, 15]. Strontia is used as a catalyst in chemical engineering [16] and is also used in cathode ray tubes for the blocking of X-ray emission.

Knowledge of bandgap engineering and optical properties of semiconducting compounds is a promising field of study for the design of optoelectronic and photonic devices. A systematic change in the bandgap of a binary material can be achieved by the partial substitution of anion or cation of a crystal by an element of desirable properties.

In the present theoretical work bandgap of rock salt CdO is varied systematically by alloying with Sr. In order to investigate optoelectronic nature of these alloys their structural, electronic and optical properties are calculated. All calculations are based on density functional full potential linearized augmented plane wave (FP-LAPW) method with Wu-Cohen generalized gradient approximation (GGA).

## II. Theory and calculation

The calculations are performed by FP-LAPW approach within the framework of density-functional theory (DFT). The exchange-correlation energy of electrons is described by GGA. The GGA functional form Wu-Cohen is used within the wien2k package [17].
In the full potential scheme both, the potential and charge density are expended into two different basis. Within the atomic sphere the potential is expanded in spherical harmonics while outside the sphere (interstial region) it is expanded in plane wave basis. The expanded potentials are:

$$V(r) = \begin{cases} \sum_{lm} V_{lm}(r) Y_{lm}(r) & \text{...........(a)} \\ \sum_{k} V_k e^{ikr} & \text{...........(b)} \end{cases} \quad (1)$$

Inside the sphere potential is spherically symmetric while outside the sphere it is constant. $R_{MT}$ is chosen in such a way that there is no charge leakage from the core and hence total energy convergence is ensured. $R_{MT}$ values of 2.41, 2.33 and 2.19 a.u. are used for Cd, Sr and O respectively. For wave function in the interstitial region the plane wave cut-off value of $K_{max} = 7/R_{MT}$ was taken. Monkhorst-Pack grids k-points from 5×5×5 upto 43×43×43 are used in the Brillouin zone integration and convergence was checked through self consistency. The convergence was ensured for less than 1 mRy/a.u.

Structural properties of $Cd_{1-x}Sr_xO$ are calculated using Murnaghan's equation of state [18]:

$$E(V) = E_0 + \frac{B_0 V}{B_0'}\left(\frac{(V_0/V)^{B_0'}}{B_0' - 1} + 1\right) - \frac{B_0 V_0}{B_0' - 1} \quad (2)$$

Where $E_0$ is the total energy of the supercell, $V_0$ is the unit volume, $B_0$ is bulk modulus at zero pressure and $B'_0$ is derivative of bulk modulus with pressure.

Optical properties of $Cd_{1-x}Sr_xO$ are calculated using a fine k mesh of 3500 points for the present calculation. The dielectric function of a crystal depends on the electronic band structure and its investigation by optical spectroscopy is a powerful tool in the determination of the overall optical behavior of a compound. It can be divided into two parts; real and imaginary:

$$\varepsilon(\omega) = \varepsilon_1(\omega) + i\varepsilon_2(\omega) \tag{3}$$

The imaginary part of the complex dielectric function, $\varepsilon_2(\omega)$, in cubic symmetry compounds can be calculated by the following relation [19, 20]:

$$\varepsilon_2(\omega) = \frac{8}{2\pi\omega^2} \sum_{nn'} \int |p_{nn'}(k)|^2 \frac{dS_k}{\nabla \omega_{nn'}(k)} \tag{4}$$

While $\varepsilon_2(\omega)$ is used to calculate the real part of the complex dielectric function [13,16]:

$$\varepsilon_1(\omega) = 1 + \frac{2}{\pi} p \int_0^\infty \frac{\omega' \varepsilon_2(\omega')}{\omega'^2 - \omega'^2} d\omega' \tag{5}$$

Refractive index is calculated in terms of real and imaginary parts of dielectric function by the following relation:

$$n(\omega) = \frac{1}{\sqrt{2}} \left[ \left\{ \varepsilon_1(\omega)^2 + \varepsilon_2(\omega)^2 \right\}^{\frac{1}{2}} + \varepsilon_1(\omega) \right]^{\frac{1}{2}} \quad (6)$$

While absorption coefficient can be calculated by [20]:

$$\alpha(\omega) = \frac{\omega \varepsilon_2(\omega)}{c} \quad (7)$$

## III. Result and Discussion

In order to study the structural properties of $Cd_{1-x}Sr_xO$ ($0 \leq x \leq 1$), alloys are modeled at various compositions of Sr with a step of 0.25. Structure optimization of each compound is performed by minimizing the total energy with respect to the unit cell volume and c/a ratio using Murnaghan's equation of state [18]. The crystal structure of CdO and SrO is rock salt with space group Fm-3m (no. 225). The substitution of Sr in CdO affects the crystal structure significantly. The space group of $Cd_{1-x}Sr_xO$ changes to Pm-3m (no 221) at x =0. 25 and 0.75, while for x = 0.50 the structure becomes tetragonal with space group P4/mmm (no 123). The change in the crystal structure is due to the difference in the ionic radii of Cd and Sr. Similar changes in the structure of $Zn_{1-x}Cr_xSe$ is reported by Ref. [21] and in $Zn_xCd_{1-x}O$ by Ref. [13, 22]. Structural parameters such as lattice constant, a and c (A°), are caculated from the stable volume and are presented in Table 1. It is clear from the table that our calculated results for the binary compounds are in good agreement with the available experimental and calculated data. It is also clear from Fig. 1 and Table 1 that the density functional calculations performed for $Cd_{1-x}Sr_xO$ at x = 0.25, 0.75 are in good agreement with the results obtained by Vegard's law. Large deviation can be seen at x = 0.50 due to the structural change from cubic to tetragonal.

The calculated band structure and density of states for $Cd_{1-x}Sr_xO$ ($0 \leq x \leq 1$) are presented in Fig. 2. It is clear from the figure that $Cd_{1-x}Sr_xO$ ($0 \leq x \leq 1$) is an indirect bandgap material. The substitution of Sr does not affect the indirect bandgap nature of the compound but increases the gap which is clear from Fig. 2(d). The indirect bandgap varies from 0.85 to 6.00 eV and the direct bandgap also increase from 2.8 to 6.2 eV with the increase in Sr concentration. It is obvious from the data presented in Table 2 that our calculated values for the bandgaps of CdO and SrO are closer to experimental results than other calculated ones. The reason for our better results is the use of effective Wu-Cohen potential in the GGA scheme [27], and high k-points (3500). The origin of band structures presented in Fig. 2 can be understood by the corresponding density of states [28]. It is clear from the figure that the conduction band is mainly composed of Sr-d state for all ternary alloys. Figs. 2(a) and 2(b) show that the lower part of the valence band is composed of Cd-4d and the upper part is mainly dominated by O-2p state. While in Fig. 2(c), it is shown that the lower part of the valence is composed O-2p and the upper part is of Cd-4d state. This interchange of states is due to the change in the nature of bonding. At x = 0.25 and 0.50 partial covalent bond is stronger so the charge is shared by O-p and Sr-s states, while at x = 0.75 the bond nature is more ionic than covalent in O-2p and Sr-3p state (the charge is transferred among the ions, Sr, O). The change in bond nature is due to the increase in electronegativity difference of cation and anion with concentration x.

The variation in the bandgap of $Cd_{1-x}Sr_xO$ ($0 \leq x \leq 1$) provides promising results of the use of the compound in optoelectronic devices working in visible to ultraviolet region. Depending upon the need and requirement of a particular application, any desired bandgap between 0.85 and 6.00 eV can be achieved.

The calculated imaginary part of the dielectric function for $Cd_{1-x}Sr_xO$ ( $0 \leq x \leq 1$ ) in the energy range 0 – 40 eV is shown in Fig. 3. It is clear from the figure that for x = 0, 0.25, 0.50, 0.75 and 1.0 the critical points in the imaginary part of the dielectric function occurs at about 2.80 eV, 2.83 eV, 3.20 eV, 4.25 eV and 5.95 eV respectively. These points are closely related to the direct bandgaps $E_g^{\Gamma-\Gamma}$; 2.80 eV, 2.85 eV, 3.30 eV, 4.30 eV and 6.20 eV of $Cd_{1-x}Sr_xO$ for the corresponding values of x = 0, 0.25, 0.50, 0.75 and 1.

The absorption of a material can be easily described by its $\varepsilon_2(\omega)$ plot. It is clear from Fig. 3 that CdO has high absorption in the region 2.8 eV to 18.5 eV with different peaks. The width and critical points of the absorption region, shifts towards higher energy as the concentration of Sr increases from 0 to 100 %. The structure of the compound changes from cubic to tetragonal when 50 % Cd is substituted by Sr and hence the material becomes anisotropic. An extraordinary peak at 7 eV in the c-direction shows strong absorption of $Cd_{0.50}Sr_{0.50}O$. It is also clear from the figure that anisotropy decreases with the increase in the incident photon energy. The material becomes isotropic for high energy photons. The variation in the absorption spectra for different concentrations of Sr can be related to the change in the bandgap. Materials with bandgap lesser than 3.1 eV work well in the visible light devices while those with bandgap larger than 3.1 eV can be used in UV devices [29, 30]. The prominent variations in the optical absorption region with bandgaps of $Cd_{1-x}Sr_xO$ (2.8 eV to 6.20 eV) confirms its suitability for optical device working in the major parts of the spectrum; visible and UV.

The calculated real parts of the complex dielectric function $\varepsilon_1(\omega)$ for $Cd_{1-x}Sr_xO$ is presented in Fig. 4. The figure clearly shows the anisotropic nature of the tetragonal structure ($Cd0_{0.50}Sr_{0.50}O$). It is clear from the figure that the static dielectric constant, $\varepsilon_1(0)$, is strongly dependent on the bandgap of the compound. The calculated values of $\varepsilon_1(0)$ for $Cd_{1-x}Sr_xO$ at x =

0, 0.25, 0.50, 0.75 and 1.0 are 3.8, 3.62, 3.30, 2.95 and 2.64 for corresponding direct bandgaps 2.80 eV, 2.85 eV, 3.30 eV, 4.30 eV and 6.20 eV respectively. This data explains that the smaller energy gap yields larger $\varepsilon_1(0)$ value. This inverse relation of $\varepsilon_1(0)$ with the bandgap can be explained by Penn Model [31].

$$\varepsilon(0) \approx 1 + (\hbar\omega_p / E_g)^2 \qquad (8)$$

The above relation can be used to calculate $E_g$ using values of $\varepsilon_1(0)$ and plasma energy $\hbar\omega_p$. In CdO the main peak is at 2.99 eV and it shifts towards higher energies as the Sr concentration increase in the $Cd_{1-x}Sr_xO$ crystal. The maximum shift of the peak is 6.17 eV for SrO.

The calculated refractive index for $Cd_{1-x}Sr_xO$ at x = 0, 0.25, 0.50, 0.75 and 1.0 are plotted in Fig. 5. A broad spectrum of $n(\omega)$ over a wide energy range is noted for these compounds. It is clear from the figure that the refractive index of the material decreases with the increase in the Sr concentration. For $Cd_{50}Sr_{50}O$ the structure is tetragonal which causes anisotropy and produces birefringence in the material. Extra-ordinary refractive index ($n^\perp$) for $Cd_{0.50}Sr_{0.50}O$ is larger than the ordinary refractive index ($n^{||}$). The lager value of $n^\perp$ conforms that the material is positive birefringence crystal.

Fig. 5 shows three different features of the refractive index of $Cd_{1-x}Sr_xO$. First; a maxima can be observed in the form of a hump in the spectrum at a particular energies and, second; the maxima shifts to higher energy region with the increase in Sr concentration. The third important point is that the refractive index drops below unity at certain energy ranges. For a refractive index lesser than unity *($v_g$=c/n)* means that the group velocity of the wave packet is larger than c. In other words group velocity shifts to negative domain and the medium shifts from linear to non-linear and hence the material becomes superluminal for high energy incident photons [32, 33].

## IV. Conclusion

Density functional calculations are carried out for the first time to investigate structural and optoelectronic properties $Cd_{1-x}Sr_xO$ ($0 \leq x \leq 1$). Structure as well as bonding nature of the material significantly varies with Sr concentration. The lattice constant of the crystal increases linearly with $x$, except for $x=0.5$ because of its tetragonal rather than cubic structure.

The calculated band structure predicts that the alloys have indirect bandgaps and it increases with the increase in x. On the basis of wide range of fundamental indirect bandgaps (0.85 eV to 6.00 eV) and direct bandgaps (2.8 eV to 6.2 eV), it can be conclude that the material can be used in optoelectronic devices working in the IR, visible and UV region of spectrum.

# References

[1] M. Joseph, H. Tabata, and T. Kawai, Jpn J. Appl. Phys. Part 1 **38,** L1205 (1999).

[2] M. Venkatesan, C. B. Fitzgerald, J. G. Lunney, and J. M. D. Coey, Phys. Rev. Lett. **93** 177206 (2004).

[3] C. Klingshirn, H. Priller, M. Decker, J. Bruckner, H. Kalt, R. Hauschild, J. Zeller, A. Waag, A. Bakin, H. Wehmann, K. Thonke, R. Sauer, R. Kling, F. Reuss, and C. Kirchner, Adv. Solid State Phys. **45,** 275 (2005).

[4] U. Ozgur, Y. I. Alivov, C. Liu, A. Teke, M. A. Reshchikov, S. Dogan, V. Avrutin, S. J. Cho, and H. Morkoc, J. Appl. Phys. **98,** 041301 (2005).

[5] H. J. Freund, Surf. Sci. **601**, 1438 (2007).

[6] H. Karami, A. Aminifar, H. Tavallali, and Z. Namdar, Clust. Sci. **21,** 1 (2010).

[7] R. A. Ismail, J. Mater Sci: Mater Electron, **20,** 1219 (2009).

[8] T. Ma, and D. Shim, Thin Solid Films **410**, 8 (2002).

[9] O. G. Daza, A. C. Readigos, J. Campos, M. T. S. Nair, and P. K. Nair Mod Phys Lett B **17**, 609 (2001).

[10] M. Yan, M. Lane, C. R. Kannewurf, and R. P. H. Chang Appl Phys Lett. **78,** 02342 (2001).

[11] A. A. Dakhel, and A. Y. Ali-Mohamed, J Sol-Gel Sci Technol, **44,** 241 (2007).

[12] O. Madelung, Semiconductors: Data Handbook, 3rd edn. (Springer, Berlin, 2004).

[13] H. Rozalea, B. Bouhafsa, and P. Ruteranab, Superlattices and Microstructures **42,** 165 (2007).

[14] A. S. Rao, and R. J. Kearney Phys. Status Solidi b **95,** 243 (1979).

[15] Y. Duan, L. Qin, G. Tang, and L. Shi, Eur. Phys. J. B **66,** 201 (2008).

[16]   M. Labidi, S. Labidi, S. Ghemid, H. Meradji, and F. E. H. Hassan, Phys. Scr. **82,** 045605 (2010).

[17]   P. Bhala, K. Schwarz, G.K.H. Madsen, D. Kvanicka, J. Luitz, 2001 WIEN2K, An Augmented Plane Wave+Local Orbital Program for Calculating Crystal Properties Karlheinz Schwarz, Techn. Universitat, Wien, Austria.

[18]   F. D. Murnaghan, Proc. Natl. Acad. Sci. USA **30,** 244 (1944).

[19]   F. Wooten, Optical properties of Solids, Academic Press, New York, 1972.

[20]   M. Fox, Optical Properties of Solids, Oxford University Press, 2001.

[21]   X. Ge, and Y. Zhang, Journal of Magnetism and Magnetic Materials, **321**, 198 (2009).

[22]   T. Makino, Y. Segawa, M. Kawasaki, A. Ohtomo, R. Shiroki, K. Tamura, T. Yasuda, and H. Koinuma, Appl. Phys. Lett. **78** 1237 (2001).

[23]   A. Schleif, C. Rodl, F. Fuchs, J. Furthmuller, and F. Bechstedt, Phy review B, **80**, 035112 (2009).

[24]   G. Choudhary, V. Raykar, S. Tiwari, A. Dashora, and B. L. Ahuja, Phys. Status Solidi B, 1 (2010).

[25]   M. A. Kanjwal, N. A. M. Barakat, Faheem A. Sheikh, and H. Y. Kim, J. Mater. Sci. **45,** 1272 (2010).

[26]    R. J. Zollweg *Phys. Rev.* **111,** 113 (1958).

[27]   Z. Wu, R. E. Cohen, Phys. Rev. B. **37,** 235116 (2006).

[28]   M. Maqbool, B. Amin, and I. Ahmad, JOSA B **26**, 2180 (2009).

[29]   T. H. Gfroerer, L. P. Priestley, F. E. Weindruch, and M. W. Wanless, Appl. Phys. Lett. **80**, 4570 (2002).

[30]   M. L. Benkhedir, M. S. Aida, A. Stesmans, and G. J. Adriaenssens, J. Optoelectron. Adv. Mater. **7,** 329 (2005).

[31]   D. Penn, Phys. Rev. **128**, 2093 (1962).

[32]   L. J. Wang, A. Kuzmich, and A. Dogariu, Nature **406**, 277 (2000).

[33]   D. Mognai, A. Ranfagni, and R. Ruggeri, Phys. Rev. Lett. **84**, 4830 (2000).

**Figures captions**

**Fig.1.** Variation of lattice constant (a and c) as a function of composition $x$

**Fig.2**. Calculated band structure for (a) $Cd_{0.75}Sr_{0.25}O$ (b) $Cd_{0.75}Sr_{0.25}O$ (c) $Cd_{0.25}Sr_{0.75}O$

(d) Variation of bandgap energy as a function of Sr concentration $x$

**Fig.3.** Frequency dependent imaginary part of the dielectric functions of $Cd_{1-x}Sr_xO$

**Fig.4.** Frequency dependent real part of the dielectric functions of $Cd_{1-x}Sr_xO$.

**Fig.5.** Frequency dependent refractive index of $Cd_{1-x}Sr_xO$

**Table: 1.** Lattice constants of $Cd_{1-x}Sr_xO$ ($0 \leq x \leq 1$) compared with experimental results, Vegard's law and other theoretical calculation

| | Lattice constant a and c (A°) | | | |
|---|---|---|---|---|
| x | this work | expriment | Vegard's law | other calculation |
| 0 | 4.69 | 4.689[12] | | 4.664[15] |
| 0.25 | 4.80 | | 4.79 | |
| 0.50 | 3.49, 5.56 | | 4.89 | |
| 0.75 | 5.00 | | 4.99 | |
| 1 | 5.10 | 5.159 [12] | | 5.093[15] |

**Table: 2.** Fundamental indirect and direct bandgaps of $Cd_{1-x}Sr_xO$ ($0 \leq x \leq 1$) compared with experimental and other calculation

| | Fundamental Energy bandgape (eV) | | | Direct Energy bandgap $E_g^{\Gamma-\Gamma}$ | | |
|---|---|---|---|---|---|---|
| x | this work | expriment | other calculation | this work | expriment | other calculation |
| 0 | 0.85 | 1.09[24] | 0.68[23] | 2.80 | 2.5[25] | 1.81[23] |
| 0.25 | 1.00 | | | 2.85 | | |
| 0.50 | 1.60 | | | 3.30 | | |
| 0.75 | 2.40 | | | 4.30 | | |
| 1 | 6.00 | 5.71[14] | 3.078[15] | 6.20 | 5.90[26] | 4.256[15] |

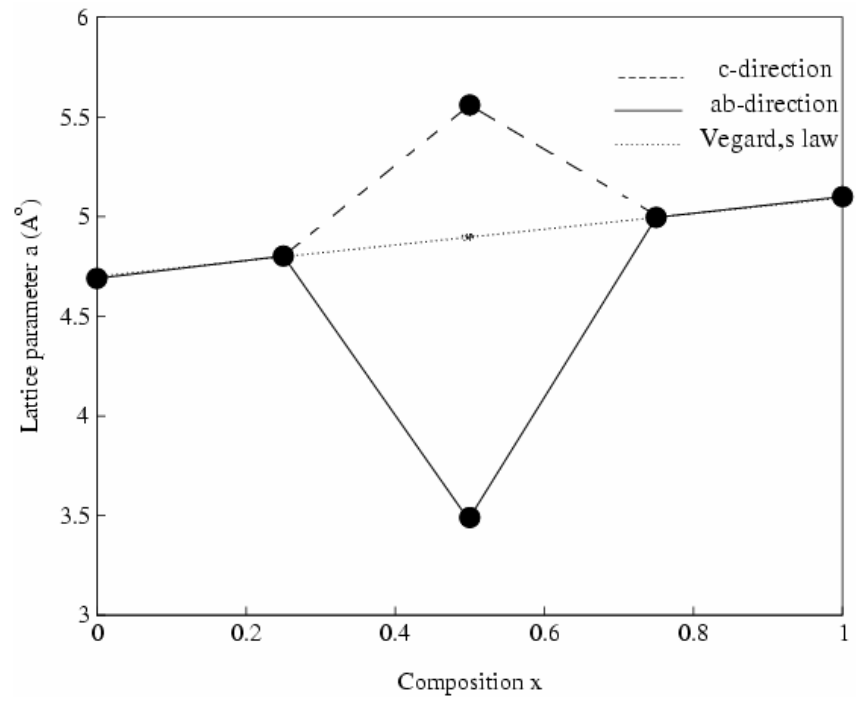

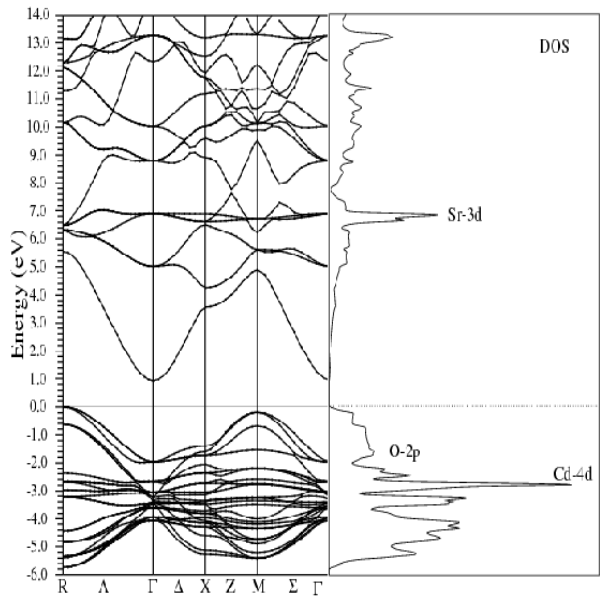

(a)

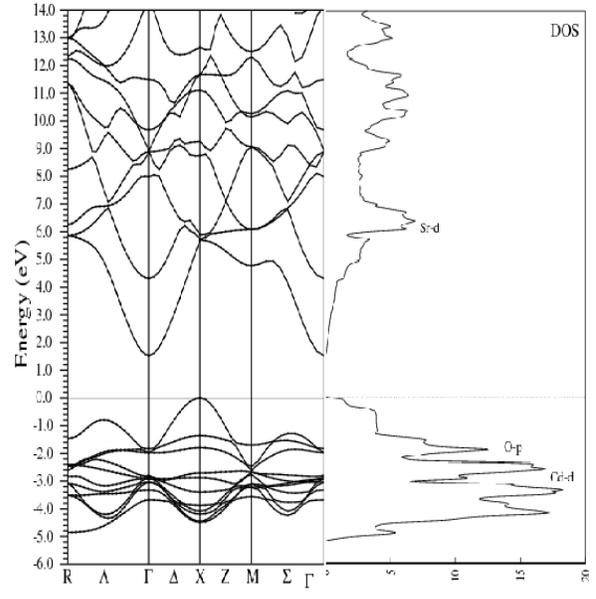

(b)

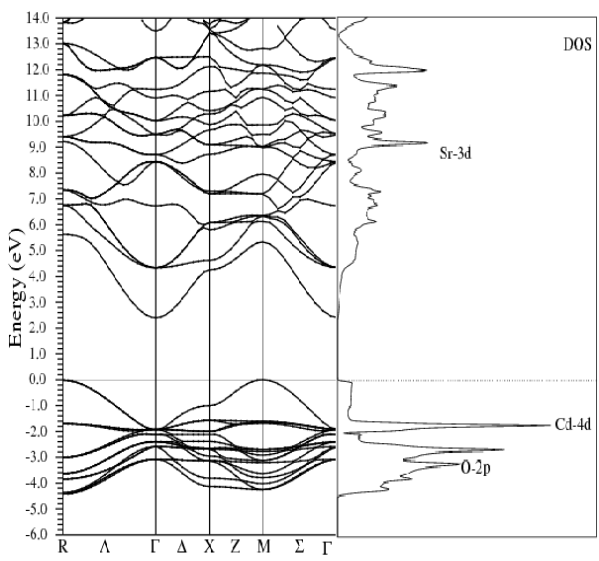

(c)

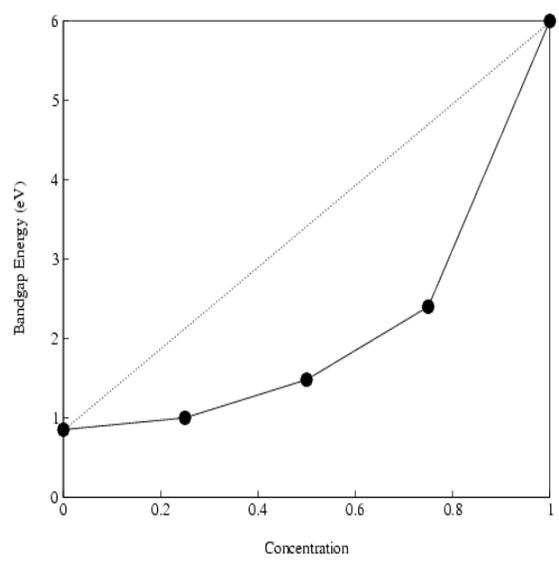

(d)

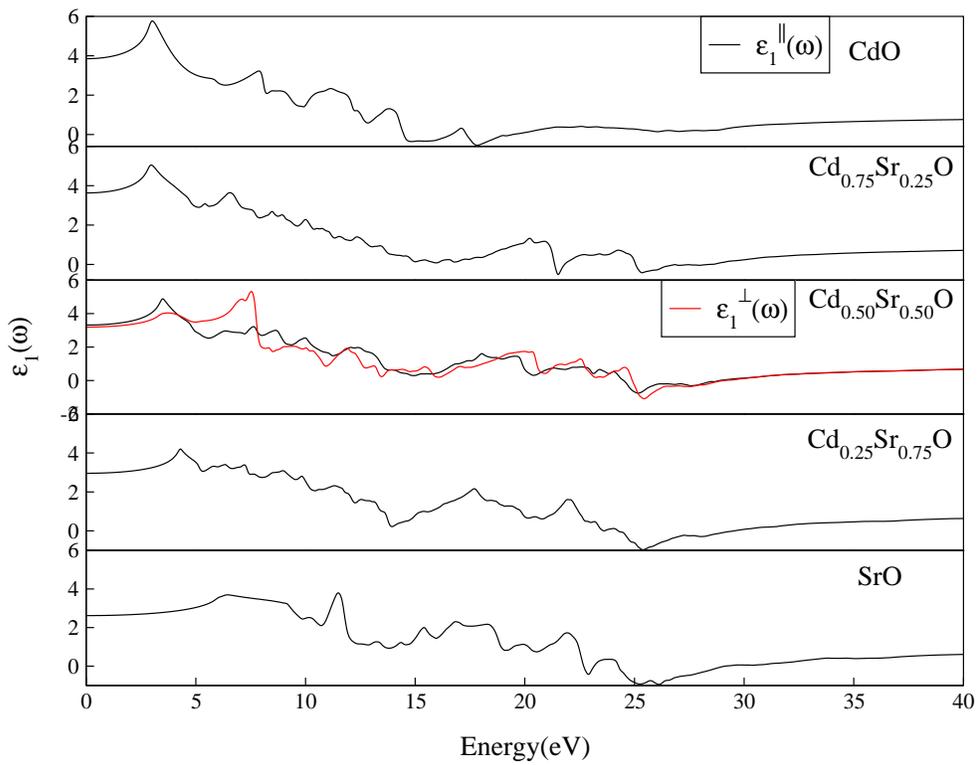

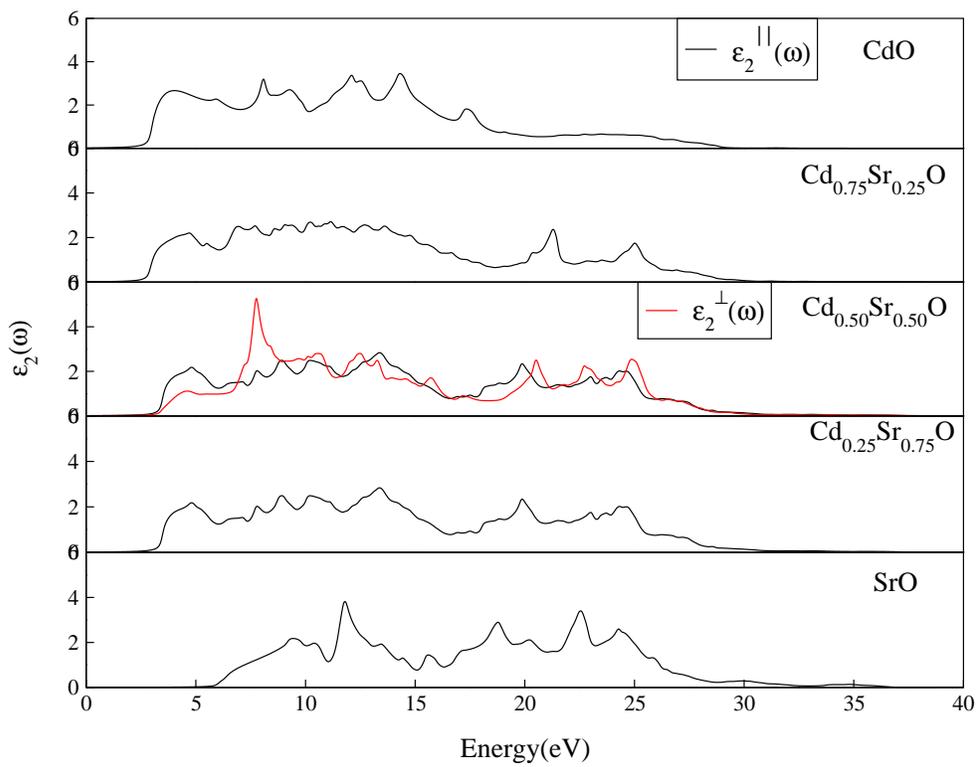

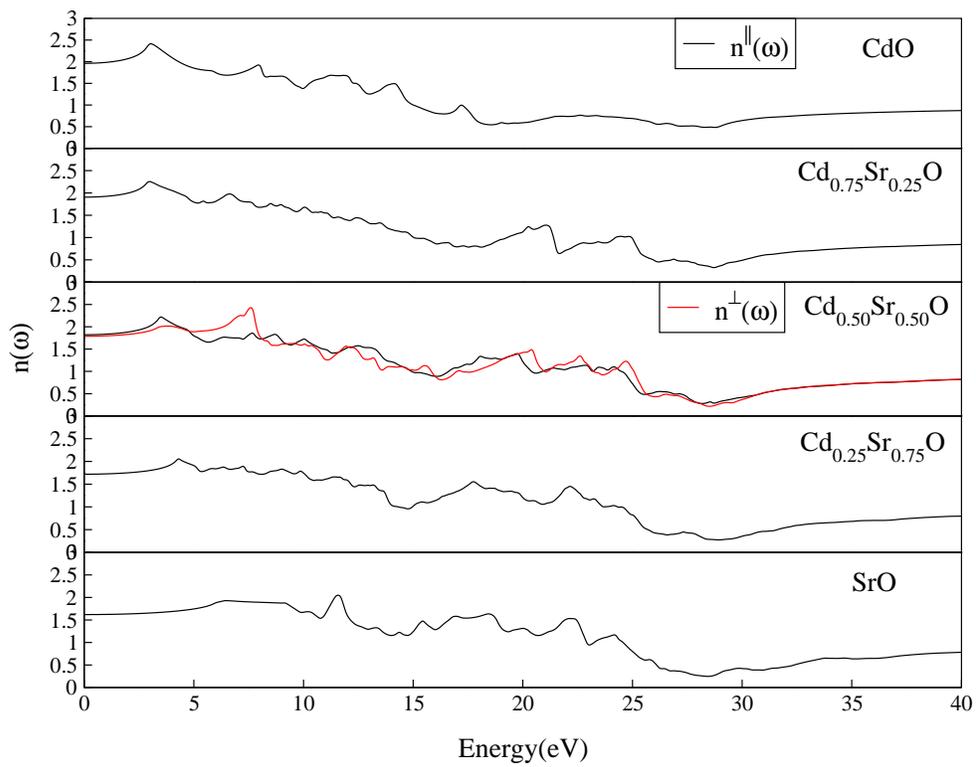